\documentclass[12pt]{article}

\usepackage{epsfig}
\usepackage{cite}
\begin{document}
\newcommand{\be}{\begin{equation}}
\newcommand{\ee}{\end{equation}}

\newcommand{\cc}{\cite}
\newcommand{\ba}{\begin{eqnarray}}
\newcommand{\ea}{\end{eqnarray}}
\begin{titlepage}
\begin{center}

  \Large  Anomalous Quark Chromomagnetic Moment and Dynamics of Elastic Scattering
  \vspace{0.50cm}\\
  Nikolai Kochelev\footnote{kochelev@theor.jinr.ru}  and  Nikolai Korchagin\footnote{korchagin@theor.jinr.ru}\\
\vspace{0.5cm}
   {\small \it Bogoliubov Laboratory of Theoretical
   Physics,\vspace{0.2cm}
Institute for Nuclear Research,\\ Dubna, Moscow region, 141980
Russia} \vskip 1ex
\end{center}

\vskip 0.5cm \centerline{\bf Abstract} We estimate the
contribution of nonperturbative quark-gluon chromomagnetic
interaction to the high energy elastic proton-proton cross section
at large momentum transfer. It is shown that this contribution is
very large in the accessible kinematic region  of the present
experiments.
 We argue that  Odderon which is the $P=C=-1$ partner of
Pomeron, is governed by the spin-flip component related to the
nonperturbative three-gluon exchange induced by the anomalous
quark chromomagnetic moment. We discuss the possible spin effects
in the elastic proton-proton and proton-antiproton scattering
coming from the interference  of spin-flip nonperturbative Odderon
and nonspin-flip Pomeron exchanges.

 \vspace{1cm}
\end{titlepage}

\setcounter{footnote}{0}

\section{Introduction}

High energy elastic proton-proton and proton-antiproton cross
sections reveal very complicated dynamics which is rather
difficult to explain within the framework of Quantum
Chromodynamics (QCD) (see the discussion in
\cite{Krisch:2010hr,Fiore:2008tp,Dremin:2013xza,
Uzhinsky:2011qu,Bourrely:2012hp,Selyugin:2013cia,Martynov:2013ana,Donnachie:2013xia,Khoze:2013jsa}).
In a conventional approach at small transfer momentum experimental
data can be described quite well by the diffractive scattering
induced by Pomeron exchange between hadrons. At large $-t \gg 1$
GeV$^2$ in the popular Donnachie-Landshoff (DL) model the dominant
contribution comes from the exchange by Odderon which is the
$P=C=-1$ partner of Pomeron. It was suggested that this effective
exchange originated from the perturbative three gluon exchange in
the proton-proton and proton-antiproton scattering
\cite{Donnachie:1979yu}.
  The experimental support for the existence  of such exchange comes
from high energy ISR data on the difference in the dip structure
around $\mid t\mid \approx 1.4 $ GeV$^2$ in the proton-proton and
proton-antiproton differential cross sections at $\sqrt{s}=53 $
GeV \cite{landshoff}.
 However, there is no any signal for  Odderon at very small transfer
momentum. We would like to emphasize that one cannot expect the
perturbative QCD DL approach to be valid even at the  largest
transfer momentum $-t\sim 14$ GeV$^2$   accessible at ISR
energies. This is related to the fact that in the three-gluon
exchange model, which is applied   to describe elastic cross
sections in the interval $-t= 3-14 $ GeV$^2$, the average
virtuality of exchanged gluons $\hat{t}\approx t/9 $ is quite
small -$\hat{t}=0.3-1.6 $ GeV$^2$. Therefore, in this kinematic
region  nonperturbative QCD effects should be taken into account.

The attempt to include some of the nonperturbative effects into
the DL model was made in \cite{landshoff2}. In that paper  the
strength of three-gluon exchange with perturbative quark-gluon
vertices was considered as a free parameter and its value was
found from the fit of the data. Therefore, a good description of
the large $-t$ cross sections in the paper  is not the result of
calculation but rather of the fine tuning to  experimental data.

One of the successful models of  nonperturbative effects is the
instanton liquid model for QCD vacuum \cite{shuryak,diakonov}.
Instantons describe nontrivial topological gluon field excitations
in vacuum and their existence leads to the spontaneous chiral
symmetry breaking in QCD. One of the manifestations of this
phenomenon is the appearance of dynamical quark mass and
nonperturbative helicity-flip quark-gluon  interaction
\cite{Kochelev:1996pv,diakonov}. Such new interaction can be
treated as a nonperturbative anomalous quark chromomagnetic moment
(AQCM). It was shown that AQCM gives a very important contribution
the to quark-quark scattering at large energies for both polarized
and nonpolarized cases
\cite{Kochelev:1996pv,diakonov,Kochelev:2013zoa,Kochelev:2009rf,Kochelev:2006ny}.
One of the applications of these results is a new model for
Pomeron based on AQCM  and nonperturbative two gluon exchange
between hadrons suggested in \cite{diakonov,Kochelev:2009rf}.

In this paper, we extend this model to the case  of the three
gluon colorless exchange between nucleons.  It will be shown that
a nonperturbative version of the Donnachie-Landshoff Odderon model
based on AQCM describes  well high energy data for the elastic
proton-proton, proton-antiproton cross sections at large transfer
momentum. The spin effects in elastic scattering are also under
discussion.

\section{Anomalous quark chromomagnetic moment and Odderon exchange}

The interaction vertex of a massive quark with a gluon  can be
written in the following form:
\begin{equation}
V_\mu(k_1^2,k_2^2,q^2)t^a = -g_st^a[ \gamma_\mu
F_1(k_1^2,k_2^2,q^2) -
\frac{\sigma_{\mu\nu}q_\nu}{2M_q}F_2(k_1^2,k_2^2,q^2)],
 \label{vertex}
 \end{equation}
where the form factors $F_{1,2}$ describe nonlocality of the
interaction, $k_{1,2}$ is the momentum of incoming and outgoing
quarks, respectively,  $ q=k_1-k_2$, $M_q$ is the quark mass, and
$\sigma_{\mu\nu}=(\gamma_\mu \gamma_\nu-\gamma_\nu \gamma_\mu)/2$.
 Within the instanton
model  the shape  of the form factor $F_2(k_1^2,k_2^2,q^2)$ is
 \be
 F_2(k_1^2,k_2^2,q^2) =\mu_a
\Phi_q(\mid k_1\mid\rho/2)\Phi_q(\mid k_2\mid\rho/2)F_g(\mid
q\mid\rho) \ , \label{form1} \ee where \ba
\Phi_q(z)&=&-z\frac{d}{dz}(I_0(z)K_0(z)-I_1(z)K_1(z)), \nonumber\\
F_g(z)&=&\frac{4}{z^2}-2K_2(z) \label{form3} \ea are the
Fourier-transformed quark zero-mode and instanton fields,
respectively,  $I_{\nu}(z)$ and $K_{\nu}(z)$ are the modified
Bessel functions and $\rho$ is the instanton size.

AQCM is defined by formula
\begin{equation}
\mu_a=F_2(0,0,0). \nonumber
\end{equation}

For our estimation below we will use the value of AQCM $\mu_a=-1$
which is within the interval $-\mu_a\sim 0.4 -1.6$ given by the
instanton model \cite{Kochelev:2009rf}. This value  is also
supported by hadron spectroscopy (see \cite{Ebert:2009ub} and
references therein). Recently, a similar value of AQCM was also
obtained within the Dyson-Schwinger equation approach with
nonperturbative quark and gluon propagators \cite{Cloet:2013jya}.
 In Fig.1, the  Donnachie-Landshoff perturbative QCD (pQCD)
and nonperturbative  AQCM-induced three gluon exchange between two
nucleons are presented.

\begin{figure}[htb]
\includegraphics[scale=1.0]{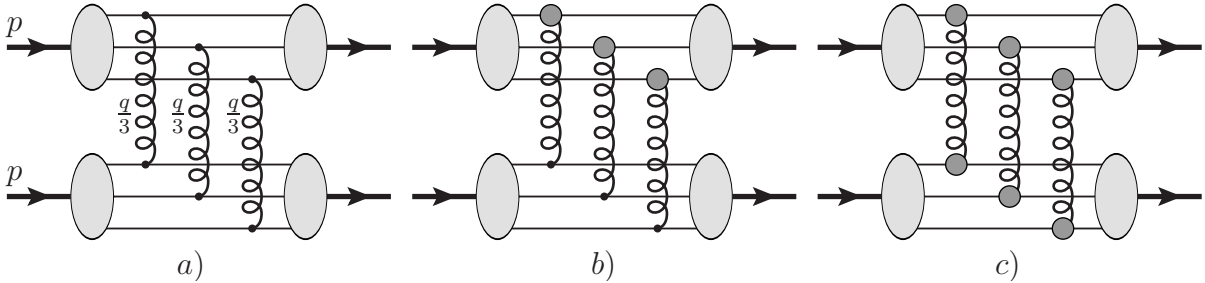}
\caption{ The left panel is the Donnachie-Landshoff mechanism for
the large $-t$ proton-proton scattering. The  right panels are the
example of  the AQCM contribution induced by the second term in
Eq.\ref{vertex}.}
\end{figure}
Within the DL model the differential cross-section of the
proton-proton and proton-antiproton scattering is given by the
formula
\begin{equation}
\frac{d\sigma}{dt}\approx \frac{244P^4}{s^6t^2R^{12}}\mid
M_{qq}(\theta)\mid^6
 \label{sigma}
\end{equation}
where  $M_{qq}$ is the matrix element at the quark level, $\theta$
is the scattering angle in the center of mass,  $P$ is the
probability of the three quark configuration in a proton, and $R$
is the proton radius. In the pQCD DL approach at the quark level
\begin{equation}
\mid
M^{pQCD}_{qq}(\theta)\mid^2=\frac{128\pi^2\alpha_s^2}{9}\frac{\hat{s}^2}{\hat{t}^2},
\label{pQCD}
\end{equation}
 where  $\hat{s}\approx s/9$,  at $\hat{s}\gg-\hat{t}$\ \ $\hat{t}/\hat{s} \sim -\sin^2\theta/4$,
and  the following  values of the parameters were taken  {\it ad hoc} :\\
\begin{equation}
P=1/10,\ \  \alpha_s=0.3,\ \  R=0.3 fm.
\end{equation}
We should emphasize that  DL assumed a very small proton radius
which is far away from the real  proton size $R\approx 1$ fm.
 For more suitable values $P=1$ and $R=1$ fm, we  got
$d\sigma/dt\sim 8\cdot 10^{-4}/t^8$ mb/GeV$^2$. It is about two
orders of magnitude less than high energy data $d\sigma/dt\approx
9\cdot 10^{-2}/t^8$ mb/GeV$^2$  at large $-t$, Fig.2 .
\begin{figure}[h] \vspace*{-0.0cm}
\centerline{\epsfig{file=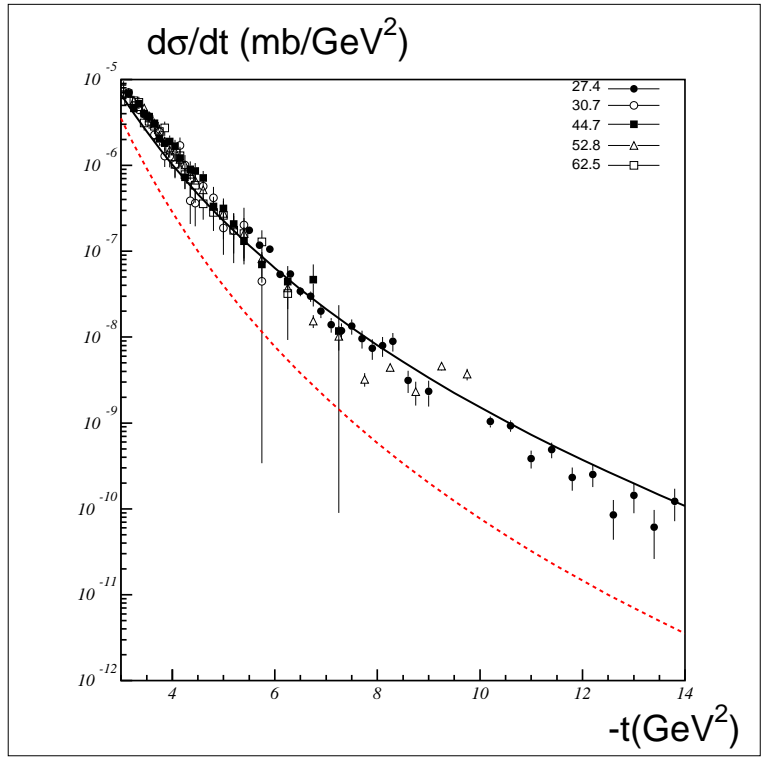,width=10cm,height=10cm,angle=0}}\
\caption{ \small The contribution of pQCD exchange (dashed line)
and AQCM   contribution (solid line)  to the elastic proton-proton
scattering at large energy and large momentum transfer in
comparison with data \cite{Nagy:1978iw}.}
\end{figure}
For the AQCM contribution at the quark level we have
\begin{eqnarray}
\mid M^{AQCM}_{qq}(\hat{s},\hat{t})\mid^2&=&\frac{16\pi^3}{3}
\alpha_s (\mid \hat t\mid) \mid\mu_a\mid
 \rho_c^2 F_g^2(\sqrt{\mid \hat{t}\mid}\rho_c)\frac{\hat{s}^2}{\mid\hat{t}\mid}
+\frac{\pi^4}{2}\mu_a^2\rho_c^4 F_g^4(\sqrt{\mid
\hat{t}\mid}\rho_c)\hat{s}^2. \label{AQCM2}
\end{eqnarray}
For estimation, we   use  $R=1$ fm, $P=1$\footnote{The value of
the strong proton radius $R\approx 1$ fm is related to the
confinement scale. The probability of the three quark
configuration in the proton $P=1$ is a natural assumption in our
three quarks on three quarks scattering model for large $-t$.},
 dynamical quark mass
$M_q=280$ MeV, average instanton size $\rho_c =1/3$ fm and the
strong coupling constant
\begin{equation}
\alpha_s(q^2)=\frac{4\pi}{9\ln((q^2+m_g^2)/\Lambda_{QCD}^2)},
\end{equation}
with $\Lambda_{QCD}=0.280$ GeV  and $m_g=0.88$ GeV
\cite{Kochelev:2009rf}. To get Eq.\ref{AQCM2} the approximation
$F_1(k_1^2,k_2^2,q^2)\approx 1$ was used and we neglected nonzero
virtuality of quarks in a proton.
 The final result for the AQCM
contribution to the proton-proton and proton-antiproton cross
section is presented by the solid line in Fig.2. We should mention
 that the AQCM contribution asymptotically decays as $1/t^{11}$ due
to the form factor Eq.\ref{form3}. Therefore, asymptotically at
very large transfer momentum  perturbative $1/t^8$ should give the
dominating contribution. However,  in the kinematic region
accessible at the present time in experiments  $-t \leq 14$
GeV$^2$, the nonperturbative AQCM contribution
 describes the available large
$-t$ data very well, Fig.2.
\begin{figure}[htb]
\centerline{\epsfig{file=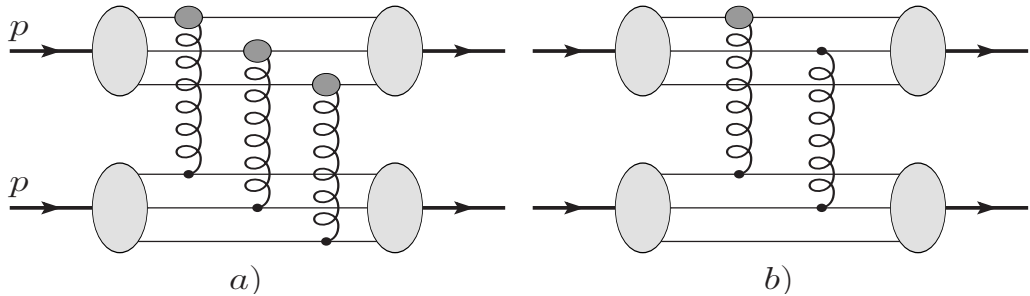,width=14cm,height=4cm,angle=0}}\
\caption{ The interference between a) DL-type AQCM  diagram and b)
Pomeron spin-flip induced by AQCM.}
\end{figure}
Finally,  some part of the difference between the structure of the
dip around $-t\approx 1-2 $ GeV$^2$ in the proton-proton and
proton-antiproton elastic scatterating at ISR energies might be
related to the difference in the sign of the interference between
the AQCM Odderon and Pomeron spin-flip amplitudes, Fig.3.

In our approach the  spin-flip component, which is proportional to
$t$, gives the dominating contribution to the negative charge
parity Odderon amplitude. In the region of small transfer momentum
this contribution to the amplitude of the  $PP$ and $P\bar P$
scattering has the dependence
\begin{equation}
 M\sim \frac{\sqrt{-t}}{(m_g^2-t)^3}, \label{small}
\end{equation}
due to  quark spin-flip induced by AQCM. In Eq.\ref{small},
$m_g\approx 0.4$ GeV is the dynamical gluon mass
\cite{Aguilar:2013hoa}.
 Therefore, the difference in the $PP$
and $P\bar P$ differential cross sections at small $-t$ and the
difference in the total  $PP$ and $P\bar P$ cross sections should
be very small at high energies. This is in  agreement with
experimental data.

Of course, one can describe  $PP$ and $\bar PP$ large $-t> 3.5 $
GeV$^2$ data  by using the assumption about a specific $t$
dependence of the Pomeron trajectory (see, for example
\cite{Bugrii:1979zh}). However, in anyway, it is necessary   to
introduce the additional $C=-1$ exchange with high intercept to
describe the difference in the $PP$ and $\bar PP$ elastic cross
sections at $\sqrt{s}=53$ GeV. A natural candidate for such
exchange is the nonperturbative three gluon DL-type  exchange. We
would like to mention that  the sizable contribution from the
conventional Pomeron exchange at large $-t>3.5$ GeV$^2$ is not
expected due to the huge suppression factor at large energies,
$(s/s_0)^{2\alpha^\prime_Pt}$,  which comes from the nonzero slope
of the Pomeron trajectory $\alpha^\prime_P\approx 0.25$
GeV$^{-2}$.

In the estimation above we assume, as in the DL model, that
momenta of exchanged gluons  are approximately equal. The
justification of this assumption   is  quite clear. To keep a
proton as a bound state of three quarks at large transfer
momentum, all quarks in the proton should scatter approximately to
the same angle. In fact, one  can also consider more complicated
multigluon contributions to elastic scattering, but we believe
that such contribution will be suppressed  by either additional
factors $\alpha_s$ or by extra factors $1/t^n$ coming from gluon
propagators and/or from form factors in the quark-gluon vertices.

\section{
 Single-spin asymmetry $A_N$ in $PP$ and $P\bar P$ elastic
 scattering}

One of the long-standing problems of QCD is the  understanding of
the large spin effects observed in the different high energy
reactions \cite{Krisch:2010hr}, \cite{Leader:2001gr}. Recently, we
have shown that the AQCM contribution leads to  very large
single-spin asymmetry (SSA) in the quark-quark scattering
\cite{Kochelev:2013zoa} and, therefore, it can be considered as a
fundamental mechanism for explanation of anomalously large SSA
observed in different inclusive and exclusive reactions at high
energy. In elastic scattering, large SSA  was found in the
proton-proton scattering at AGS energies at large transfer
momentum, Fig.4.
\begin{figure}[htb] \vspace*{-0.0cm}
\centerline{\epsfig{file=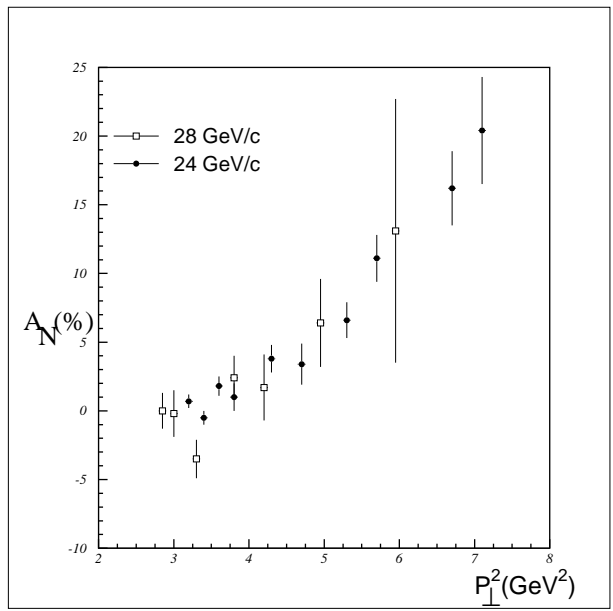,width=8cm,height=8cm,angle=0}}\
\caption{ \small Single-spin asymmetry in the elastic
$PP\rightarrow PP$ scattering at large momentum transfer at AGS
\cite{Crabb:1990as}.}
\end{figure}
In the bases of the c.m. helicity amplitudes SSA is given by the
formula
\begin{equation}
A_N=-\frac{2Im[\Phi_5^*(\Phi_1+\Phi_2+\Phi_3-\Phi_4)]}{\mid\Phi_1\mid^2+\mid\Phi_2\mid^2+\mid\Phi_3\mid^2+
\mid\Phi_4\mid^2+4\mid\Phi_5\mid^2},\label{SSA}
\end{equation}
 where the helicity amplitudes $\Phi_1=<++\mid++>$,
$\Phi_2=<++\mid-->$, $\Phi_3=<+-\mid+->$, $\Phi_4=<++\mid-->$ and
$\Phi_5=<++\mid-+>$. It is evident that due to the negative charge
parity Odderon contribution, the helicity-flip amplitude $\Phi_5$
should have a different sign for the proton-proton and
proton-antiproton scattering. Therefore, SSA in the case of the
elastic proton-antiproton scattering flips the sign in comparison
with the proton-proton scattering. This prediction can be tested
by the PAX Collaboration at HESR \cite{Barone:2005pu}. Due to the
dominance of spin one $t$-channel gluon exchanges in the structure
of Pomeron and Odderon, we can also expect that single-spin
asymmetry at large $-t$ should have a weak energy dependence. This
prediction can be checked in the polarized  proton-proton elastic
scattering in the pp2pp experiment  at RHIC in  case of extending
their kinematics to the large transfer momentum region
\cite{pp2pp} \footnote{We would like to thank Jacek Turnau for the
discussion of  this problem.}. However, the calculation of
absolute value of SSA in the elastic $PP$
 and $P\bar P$ scattering at large transfer momenta is a very difficult task, because one
 needs
 to know spin-flip and non-spin flip components of both Odderon
 and Pomeron exchanges. Furthermore, in the region of small
 transfer momenta and low energies it is needed to include the effects
 of secondary Reggion exchanges as well.

\section{Conclusion}

In summary, it is shown that the anomalous quark-gluon
nonperturbative vertex gives a large contribution to the elastic
proton-proton and proton-antiproton  scattering at large momentum
transfer. One can treat three-gluon exchange induced by this
vertex as effective Odderon exchange with the spin-flip dominance
in its amplitude.
 We should mention that the anomalous quark
chromomagnetic moment is proportional to $1/\alpha_s$
\cite{Kochelev:1996pv}. Therefore, non-spin flip component in
Odderon due to perturbative vertex should be suppressed by
$\alpha_s$ factor.
 We argue that a strong spin
dependence of the Odderon amplitude might lead to the large spin
effects in the proton-proton and proton-antiproton scattering at
large momentum transfer.

 Our approach is based on the existence of two
quite different scales in hadron physics. One of them  is related
to the confinement radius $ R\approx 1$ fm and it is consistent,
as well,  with an average distance between instanton and
antiinstanton within the  instanton liquid model, $R_{I\bar
I}\approx 1$ fm \cite{shuryak,diakonov} . This scale is
responsible for the diffractive type scattering at small momentum
transfer. Another one  is fixed by the scale of spontaneous
symmetry breaking. Within the instanton model it is given by an
average instanton size in QCD vacuum $\rho_c\approx 1/3$ fm. This
scale leads to the appearance of a large dynamical quark mass and
large anomalous quark chromomagnetic moment and is responsible for
the dynamics of the hadron-hadron elastic scattering at large
momentum transfer. We would like to mention that the two scale
model for the hadron structure was discussed in different aspects
in the papers \cite{Dorokhov:1993fc,Schweitzer:2012hh}.

\section{Acknowledgments}
The authors are very grateful to  A.E. Dorokhov, A.V. Efremov,
S.~V.~Goloskokov, E.A.~Kuraev, N.N.~Nikolaev, L.N. Lipatov, O.V.
Selyugin and V.V. Uzhinsky  for useful discussions.
 The work of N.K. was supported in part by  a visiting
scientist grant from the University of Valencia and  by the MEST
of the Korean Government (Brain Pool Program No. 121S-1-3-0318).
We also acknowledge that this work was initiated through the
series of APCTP-BLTP JINR Joint Workshops.

\end{document}